# Realization of continuously electron doping in bulk iron selenides and identification of a new superconducting zone

R. J. Sun[1,3]†, Y. Quan[2]†, S. F. Jin[1,3]*, Q. Z. Huang[4], H. Wu[4], L. Zhao[1], L. Gu[1,3]*, Z. P. Yin[2]*, X. L. Chen[1,3,5]*

[1.] *Institute of Physics, Chinese Academy of Science, Beijing, 100190, China.*

[2.] *Department of Physics and Center for Advanced Quantum Studies, Beijing Normal University, Beijing, 100875, China.*

[3.] *School of Physical Sciences, University of Chinese Academy of Sciences, Beijing, 100190, China.*

[4.] *NIST Center for Neutron Research, National Institute of Standards and Technology, Gaithersburg, Maryland 20899, USA.*

[5.] *Collaborative Innovation Center of Quantum Matter, Beijing, 100190, China.*

† R.J.S. and Y.Q. contribute equally to this work.

*Correspondence should be addressed to S.F.J. (shifengjin@iphy.ac.cn), Z.P.Y. (yinzhiping@bnu.edu.cn), L.G (l.gu@iphy.ac.cn) or X.L.C (chenx29@iphy.ac.cn).

It is known that iron selenide superconductors exhibit unique characteristics distinct from iron pnictides, especially in the electron-doped region. However, a comprehensive study of continuous carrier doping and the corresponding crystal structures of FeSe is still lacking, mainly due to the difficulties in controlling the carrier density in bulk materials. Here we report the successful synthesis of a new family of bulk $Li_x(C_3N_2H_{10})_{0.37}FeSe$, which features a continuous superconducting dome harboring Lifshitz transition within the wide range of $0.06 \leq x \leq 0.68$. We demonstrate that with electron-doping, the anion height of FeSe layers deviates linearly away from the optimized values of pnictides and pressurized FeSe. This feature leads to a new superconducting zone with unique doping dependence of the electronic structures and strong orbital-selective electronic correlation. Optimal superconductivity is achieved when the Fe 3d $t_{2g}$ orbitals have almost the same intermediate electronic correlation strength, with moderate mass enhancement between 3~4 in the two separate superconducting zones. Our results shed light on achieving unified mechanism of superconductivity in iron-based superconductors.

**Introduction**

Iron-based superconductors are the only known family of unconventional high transition temperature (Tc) superconductors besides cuprate[1,2]. Whereas pristine cuprates are Mott-insulating and superconductivity has been realized as a consequence of controlling the carrier concentration[3], the parent iron-based compounds are metallic[4], and the increase of carrier concentration as a result of doping is no longer a decisive factor for superconductivity[5]. For instance, superconductivity can be directly induced in iron-based materials by application of pressure, which mainly modified the Fe-Fe distance and the interlayer anion height within the FeAs or FeSe trilayers[6,7]. It is further suggested that during chemical doping, modification of the Fermi surface by structural distortions is more important than charge doping for inducing superconductivity, as exemplified in $Ba_{1-x}K_xFe_2As_2$ families[8]. Similarly, in the case of electron-doped FeSe - which has fascinating properties distinct from pnictides, such as the absence of hole pocket in the Fermi surface[9,10], reemergence of superconductivity under high pressure[11,12] and substantially higher Tc value in monolayer form[13,14] - the structure and properties of FeSe should also be intimately linked. However, in contrast to the well-established phase diagram for iron-pnictides[4], a comprehensive study of continuous electron doping and the corresponding crystal structure of FeSe-based superconductors is still lacking, mainly due to the difficulties in controlling the carrier density in bulk iron selenides. This, in return, prevent us to uncover the origin of the extraordinary behaviors emerged in iron selenides, and whether there are common key ingredients contributing to superconductivity in both selenides and pnictides.

Systematic control of electron doping in a bulk iron selenide superconductor is lacking. Although electron doping has been achieved in bulk FeSe-based materials such as $A_xFe_{2-y}Se_2$ (A= K, Rb, Cs and Tl/K) [15,16,17], and $(Li_{0.8}Fe_{0.2})OHFeSe$[18], the doping levels are discrete and untunable. On the other hand, by depositing K[19] or applying liquid gating[20] on the monolayer or thin flacks of FeSe, a finer control over electron doping is realized on the surface of FeSe compared to the bulk. Angle-resolved photoemission spectroscopy (ARPES) and Hall transport measurements recently recognized a superconducting dome harboring two Lifshitz transitions [21,22,23], wherein novel electronic structures and anomalous strong correlation effects have been identified. Unfortunately, no reliable structural information have been extracted from these few-layer system[24]. All these issues definitely call for a new FeSe-derived superconductor capable of continuous electron doping to unveil how the superconductivity and structure evolve in electron-doped FeSe systems.

Here, we are able to continuously dope Li into FeSe layers by co-intercalating $C_3N_2H_{10}$ molecules between the FeSe layers to form $Li_x(C_3N_2H_{10})_{0.37}FeSe$ bulk material in the whole doping region of 0.06≤x≤0.68. Based on this material, we have not only succeeded in accessing the widest doping level ever achieved in FeSe bulk materials, but also attained a continuous variation of Tc with doping above x=0.15 and an optimal Tc=46 K around x=0.37. The detailed structural variation of FeSe layers upon electron doping is found to deviate strongly from that of the FeE (E=As, P) layers in iron pnictides, and enters a new superconducting zone well separated from the known one that hosts various iron pnictides and pressurized FeSe. Our state-of-the-art density functional theory combined with dynamical mean field theory (DFT+DMFT) calculations provide a comprehensive picture on the evolution of electronic structures, orbital-selective correlation strength and spin fluctuations with electron doping in this FeSe system.

**Structural Characterization**

The phase-pure $Li_x(C_3N_2H_{10})_{0.37}FeSe$ (0.06<x<0.68) samples intercalated by 1,2-$C_3N_2H_{10}$ molecules together with the alkali metal was prepared by solvothermal method (detailed in Methods). Figure S1 shows the PXRD pattern collected for $Li_{0.26}(C_3N_2H_{10})_{0.37}FeSe$ at 295K. All Bragg peaks of the XRD pattern can be indexed using a primitive tetragonal unit cell with lattice parameters $a$= 3.8161(7) Å and $c$ = 10.8318(8) Å. The $c$ lattice parameter of the new compound is almost 100% larger than the pristine $β$-FeSe, suggesting that 1,2-$C_3N_2H_{10}$ molecules have been co-intercalated between the neighboring FeSe layers (Figure 1a). Figure 1b shows the electron charge distribution of $Li_{0.26}(C_3N_2H_{10})_{0.37}FeSe$ obtained from a Fourier analysis of the (001) diffraction lines. The charge density along the $c$ axis is consistent with the assignment of the amide molecule to the center of two adjacent FeSe layers, with the nearest-neighbor (NN) distance of 3.67 Å, in a satisfactory agreement to the NN distance in a $C_3N_2H_{10}$ molecule (3.73 Å). To locate the light scatting atoms H & Li and understand the detailed crystal structure, neutron powder diffraction (NPD) experiments were conducted with Cu311 (λ=1.5401 Å) monochromators using the BT-1 powder diffractometer at the NIST Center for Neutron Research. The relatively large background of the NPD pattern indicates large incoherent scattering from the H atoms in the sample. The crystal structure of $Li_{0.26}(C_3N_2H_{10})_{0.37}FeSe$ was solved by *ab initio* structure determination from NPD data (Figure 1c). To solve the structure, $C_3N_2H_{10}$ molecules and FeSe layers were used as independent motifs in a simulated annealing approach. A preliminary structural model is built up with space group $P42_12$ and then Li positions are located by Fourier difference analysis. The final Rietveld refinement produced an excellent fit to the diffraction pattern, with $R_p$ = 2.25% and $R_{wp}$ = 2.89%.

As shown in Figure 1(b), the intercalated $C_3N_2H_{10}$ molecules were located at 8g site and disordered in diagonal orientations. In the model, two N-H bonds of about 1 Å in the $C_3N_2H_{10}$ species are directed towards the selenide ions with H- Se distances of 2.666 Å and 2.690 Å, consistent with hydrogen bonding interactions found in the lithium/ammonia intercalates[25]. Li ions were located adjacent to the FeSe layers at sites (0, 1/2, z) (2c site), coordinated with four Se atoms with Li-Se bond lengths of 2.959 Å. Meanwhile, the Li-N bonding is also expected due to a short distance of 1.5 Å between Li and the terminal N atoms in the molecules. A plot of the observed and calculated intensities measured at 295 K is shown in Figure 1c and the refined structure parameters are listed in supplementary materials (Table S2).

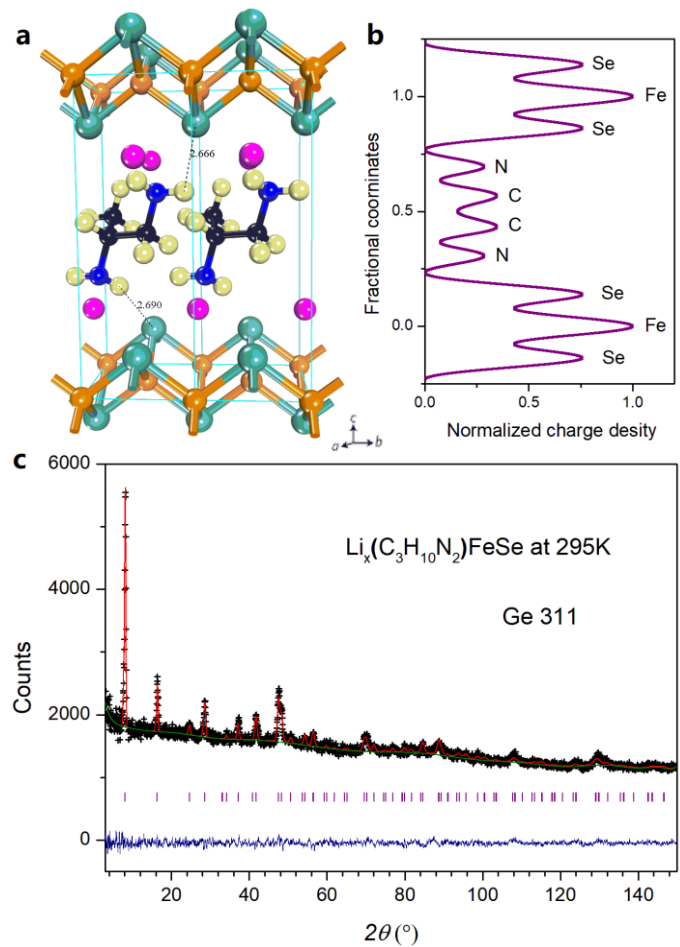

**Figure 1. Crystal structure and Rietveld refinement against NPD data for Li$_{0.26}$(C$_3$N$_2$H$_{10}$)$_{0.37}$FeSe.** **a**, A schematic view of the structure of Li$_{0.26}$(C$_3$N$_2$H$_{10}$)$_{0.37}$FeSe. In the model the anti-PbO-type FeSe layers and the C$_3$N$_2$H$_{10}$ layers stack alternately. Li atoms are found bonded to four adjacent Se atoms. **b**, Charge density along the c axis from a Fourier synthesis of the (00l) integrated intensities using the 298 K X-ray data. The peaks labeled Fe, Se, C and N represent the iron, selenide, carbon and nitrogen bounding layers, respectively. **c**, Observed (crosses) and calculated (red solid line) NPD pattern for Li$_{0.26}$(C$_3$N$_2$H$_{10}$)$_{0.37}$FeSe ( $\lambda$ =1.9395 Å) at 295 K. Differences between the observed and calculated intensities are shown at the bottom of the figure. Bragg peak positions are indicated by short purple vertical bars below the NPD patterns.

**Systematic control of the electron doping**

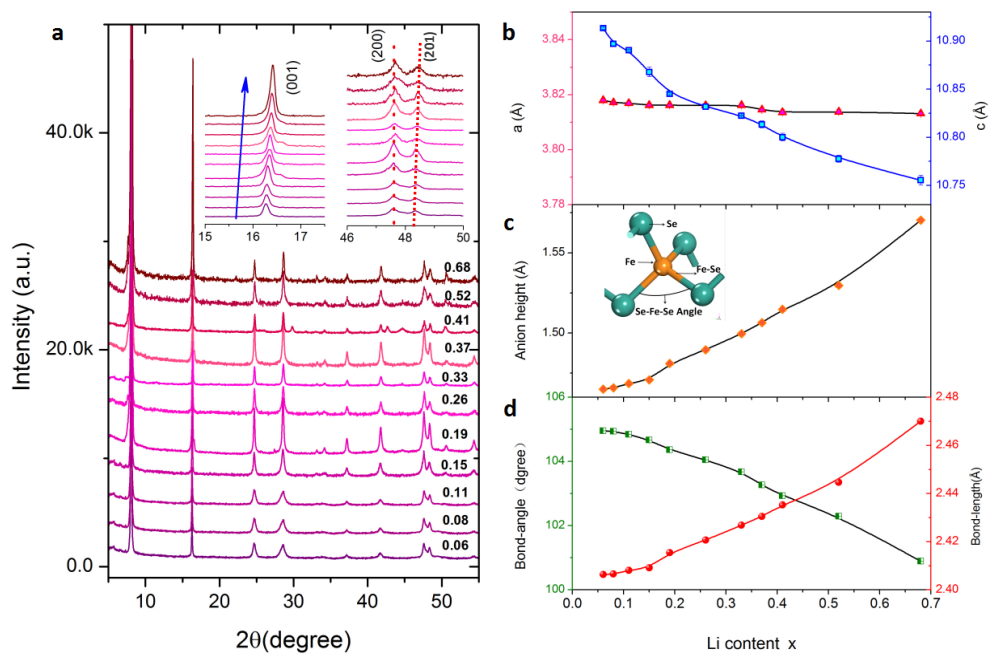

**Figure 2. Structural evolution of Li$_x$(C$_3$N$_2$H$_{10}$)$_{0.37}$FeSe as a function of Li doping level x** (a) The X-ray diffraction patterns obtained at different Li doping levels between 0.06 to 0.68. The inset displays the enlarged (002) and (200) peaks, where the shift of (002) peaks to higher angels corresponds to the lattice contraction along *c*-axis. (b) The *a* and *c* lattice constants of the tetragonal cell as a function of Li doping. There is no obvious variation in *a* while significant lattice contraction along the *c*-axis with increasing Li doping (c) (d)The Se anion height, Fe–Se bond distance and Se–Fe–Se bond angles versus Li doping. The error bars indicate one standard deviation and the inset of (c) presents the FeE4 tetrahedra model.

In contrast to the previously reported bulk compounds of Li$_{0.6}$(ND$_3$)$_{0.8}$(ND$_2$)$_{0.2}$Fe$_2$Se$_2$[25] and (Li$_{0.8}$Fe$_{0.2}$OH)Fe$_{1-y}$Se[18], in which the dopant concentrations are fixed, it is found that the Li concentration in Li$_x$(C$_3$N$_2$H$_{10}$)$_{0.37}$FeSe can be systematically tuned in a wide range (0.06<x<0.68). The success in continuously tuning the carrier concentrations by Li doping allows us to explore the evolution of the crystal structures and superconductivity with fine electron doping. The powder X-ray diffraction patterns collected for 11 samples with gradually increased Li concentrations are displayed in Figure 2a, where no new peaks beyond the P4$_2$12 structure were observed above the 1% intensity level for all the patterns. Inspection of Figure 2a and its insets immediately reveals that all the peaks related to (00l) reflections shift continuously to higher angles upon lithium insertion, while the peaks that belong to (h00) reflections remain almost invariant. The most under-doped sample without unreacted FeSe is obtained at x=0.06, with cell parameters $a$=$b$= 3.818(4) Å, $c$=10.9133(5) Å. Doping lithium gradually suppresses the $c$-axis lattice constant, while leaving the $a$ and $b$ lattice constants essentially unchanged (Figure 3b). The most heavily doped sample obtained at x=0.68 has cell parameters $a$=$b$=3.813(2) Å, $c$=10.7555(4) Å. In a previous study on the phase diagram of $(Li_{0.8}Fe_{0.2}OH)Fe_{1-y}Se$, it was found that the basal lattice parameter $a$ varies linearly with the occupancy of the Fe site in the selenide layers from ~3.75 Å to ~3.82 Å[26]. The invariant $a$-axis lattice constant in a wide range of dopant concentration is consistent with the invariant Fe occupancy across the $Li_x(C_3N_2H_{10})_{0.37}FeSe$ series by chemical analysis and the refinement from the NPD pattern (Table S2 and S3).

Figure 2 c, d summarizes the impact of Li doping on the structure of FeSe layers obtained from Rietveld refinement against the PXRD data of $Li_x(C_3N_2H_{10})_{0.37}FeSe$ series (Figure S3). Key structural parameters for iron-based superconductors are the Fe−Fe distance in the plane (=$a/\sqrt{2}$), the anion height $h_E$ (E = chalcogen or pnictogen) , the Fe−E bond length, and the E−Fe−E angles in the $FeE_4$ tetrahedra. Although the Fe-Fe distance is essentially doping independent across the series, we find the anion height increases monotonically with increasing electron doping, surpassing the $h_E$ values of all known iron pnictides in heavily electron doped region. According to our DFT+DMFT results shown below, the significantly increased $h_{Se}$ has a profound impact on the electronic structure as well as the electronic correlation. Meanwhile, for the title compounds, the $FeSe_4$ tetrahedra are extremely squashed in the basal plane relative to the more regular tetrahedra found in iron arsenides. The increased electron doping squash the $FeSe_4$ tetrahedra further, pushing the Se−Fe−Se angles from 104.95° to an extremely small value down to 100.89° in the most Li doped sample. Besides, the Fe−Se bond length is found almost linearly elongated from 2.406(3) Å to 2.470(1) Å with electron doping, which is also in contrast to the invariant Fe-E distances found in iron pnictides series with doping[27]. As the Fe–Se distance is strongly doping dependent (Figure 2d), the strong hybridization between the Fe $3d$ and the Se $4p$ orbitals is directly affected by electron doping.

**Superconductivity**

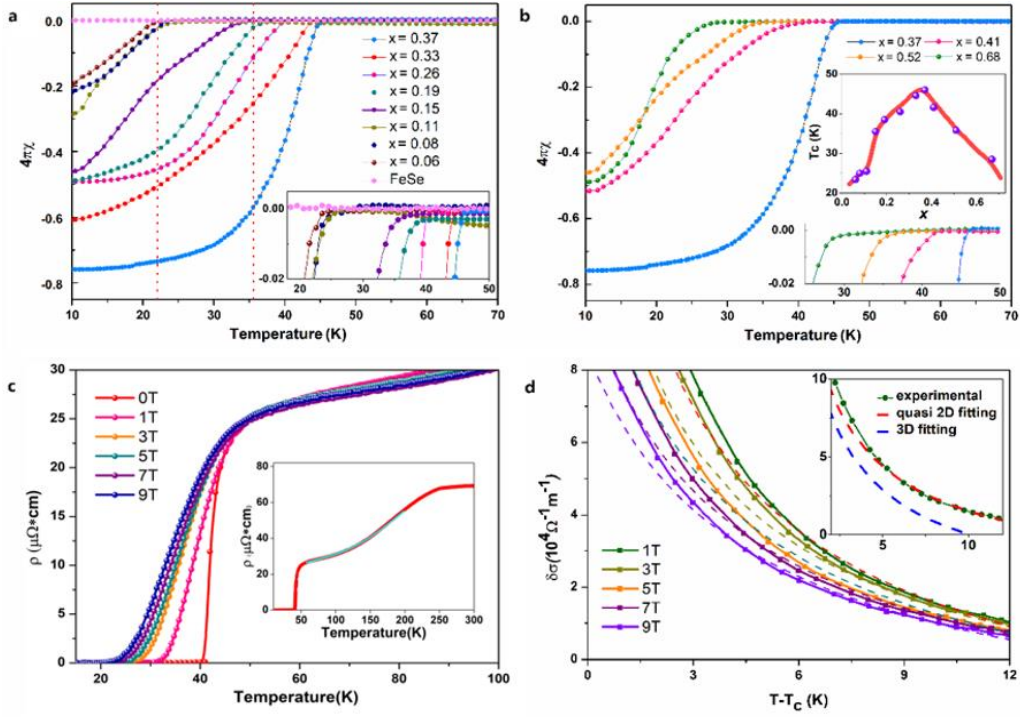

**Figure 3. Superconducting properties of Li$_x$(C$_3$N$_2$H$_{10}$)$_{0.37}$FeSe.** (a) Zero-field cooled magnetization (H$_{dc}$=10 Oe) for under-doped and optimal doped superconducting Li$_x$(C$_3$N$_2$H$_{10}$)$_{0.37}$FeSe samples with Li doping levels x= 0.06 to 0.37. (b) Magnetization data for over-doped zone with Li doping levels x= 0.37 to 0.68. Inset shows the doping dependence of the superconducting temperature. (c) Temperature dependence of the electric resistivity ρ of Li$_{0.37}$(C$_3$N$_2$H$_{10}$)$_{0.37}$FeSe under increasing magnetic fields. (d) The fluctuation contribution to the conductivity Δσ varies with temperature under external magnetic fields up to 9T. The inset is Δσ vs temperature curve and the quasi 2D and 3D fitting.

Evolution of the superconductivity with electron doping is studied through Zero-field cooled magnetization on the Li$_x$(C$_3$N$_2$H$_{10}$)$_{0.37}$FeSe series with systematically controlled Li doping. Figure 3a shows the magnetization data as a function of doping with x <0.37. For all the spectra at all doping levels, diamagnetic transition of the samples is always visible, and depends strongly on the doping level. As the Li concentration x of the samples increases from 0.06 to 0.11, the Tc gradually shifts from 22.5 K (x=0.06) to 23.6 K (x=0.11), and the superconducting volume fraction slightly increases. A new feature emerges when the Li concentration is further increased, the superconductivity is suddenly enhanced at x=0.15, with the Tc onset drastically increased to 34.5 K (Figure 3a and the inset therein). Meanwhile, a slight kink at 23.5 K can also be identified for $Li_{0.15}(C_3N_2H_{10})_{0.37}FeSe$, implying that two separate SC phases coexist in this sample. No intermediate transition temperature exists between the 11 K jump of Tc. The sudden enhancement of superconductivity is likely associated to a Lifshitz transition upon electron doping as shown below. Tc then continuously increases until x=0.37, the system reaches its optimal condition for superconductivity and the Tc peaks at 46 K (inset of Fig.3b). Further increase in Li concentration drives the system into over-doped region and leads to decreasing of both the Tc and SC volume fraction. By summarizing the superconducting transition temperature with Li concentration, we obtain a superconducting dome with the maximum Tc up to 46 K (inset of Fig. 3b), which is significantly enhanced compared to that in the stoichiometric FeSe.

Figure 3c displays the electric resistivity of $Li_{0.37}(C_3N_2H_{10})_{0.37}FeSe$. Between 55k to 200K, the ρ (T) curve shows a $T^2$-dependence in its normal state electrical resistivity (the cyan dash line), indicating a Fermi liquid behavior of the compound. As a signature of the quasi two-dimensional (2D) superconductivity in $Li_{0.37}(C_3N_2H_{10})_{0.37}FeSe$, the resistivity transition near $T_c^{onset}$ becomes too smooth to define a mean-field transition point, especially in the presence of a magnetic field. The fluctuation magnetoconductivity above the superconducting transition temperature (Tc) is hence analyzed in terms of the Lawrence-Doniach (LD) approach for 2D SC[28,29].

$$\Delta\sigma_{LD} = \frac{e^2}{64\pi\hbar}\frac{1}{h}\int_{-\pi/s}^{\pi/s} dk_z \left[\psi^1\left(\frac{\varepsilon + h + \omega_{k_z}^{LD}}{2h}\right) - \psi^1\left(\frac{c + h + \omega_{k_z}^{LD}}{2h}\right)\right]. \quad (1)$$

$$\omega_{k_z}^{LD} = r[1 - \cos(\dot{k}_z s)]/2) \quad (2)$$

Here $\psi^1$ is the first derivative of digamma function, e is electron charge, *h* is the reduced magnetic field, *ε* is ln(T/$T_c^{mid}$), *s* is the distance between adjacent FeSe layers and *r* is a fitting parameter corresponding to coherence length amplitude $\zeta(0)$. Figure 3d presents the fluctuation contribution to the conductivity Δσ which varies with temperature under external magnetic fields from 0 T to 9 T. The fittings lead to an averaged $\zeta(0)$ = 1.0815Å, a value one order of magnitude smaller than the FeSe inter-layer distance. As shown in the inset of Fig 3d, the experimental Δσ strongly deviates from the 3D anisotropic Ginzburg-Landau (GL) model, and agrees well with the 2D-Lawrence-Doniach model, clearly showing the superconductivity of $Li_{0.37}(C_3N_2H_{10})_{0.37}FeSe$ is quasi-2D in nature. The fitted data also yields a zero temperature upper critical field $H_{c2}(0)$ = 82T. Fig S4 shows the field dependence of the initial magnetization susceptibility of $Li_{0.37}(C_3N_2H_{10})_{0.37}FeSe$, in which the type-II SC is observed, with the fitted lower critical field $H_{c1}(0)$ = 0.15 T.

**Electronic structure**

Experimentally, the Se height $h_{Se}$ of FeSe layers is previously unknown at different doping levels [30]. With the detailed crystal structure at various doping levels obtained in the present $Li_x(C_3N_2H_{10})_{0.37}FeSe$ system, we carry out first-principles density functional theory combined with dynamical mean field theory (DFT+DMFT) calculations [31,32] to further understand its phase diagram and the corresponding electronic structures at various doping levels. Computational details are given in the method section and in the supplementary materials.

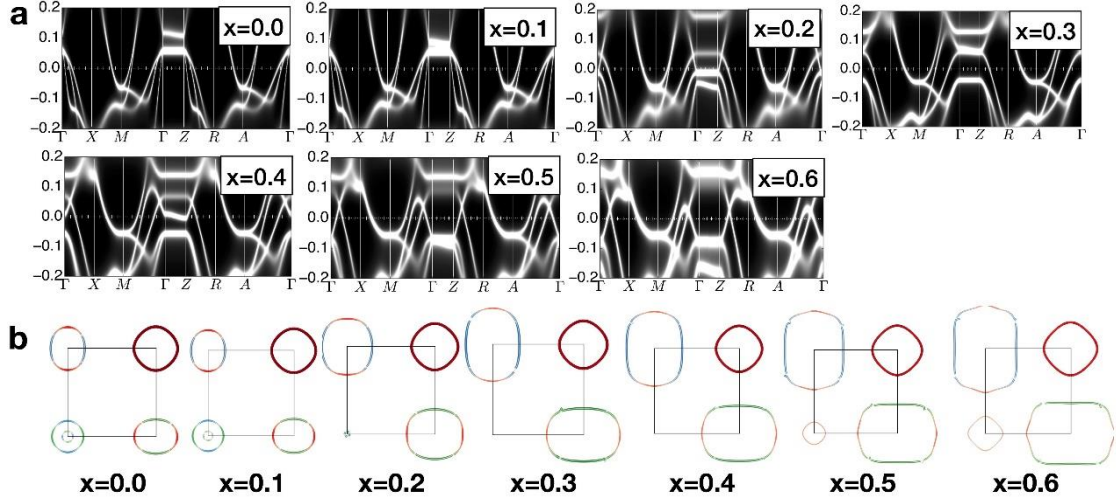

**Figure 4. a**) Momentum-resolved spectral function and **b)** the unfolded orbital resolved Fermi surfaces in the one-Fe Brillouin zone of $Li_x(C_3N_2H_{10})_{0.37}FeSe$ at different doping levels x. The blue/green color in panel **b)** denotes dominating $d_{xz}/d_{yz}$ orbital character and red color denotes dominating $d_{xy}$ orbital character. The hole pocket with dominating $d_{xy}$ orbital character persists in the whole doping range, while the other two-hole pockets with mostly $d_{xz/yz}$ orbital character gradually disappear with increasing doping level towards x=0.2, at which doping level, a new electron pocket appears at Γ. Notably, a Lifshitz transition between x=0.1 and x=0.2 occurs which is probably responsible for the experimental $T_c$ jump around x=0.15. The expansion of electron pockets at the zone corner as Li doping increases is within expectation.

Figure 4 shows the evolution of the momentum-resolved spectral function and the unfolded orbital resolved Fermi surfaces (FS) in the one-Fe Brillouin zone of $Li_x(C_3N_2H_{10})_{0.37}FeSe$ with Li doping level x. At zero doping (x=0), the spectral function resembles that of the stoichiometric FeSe. The three-hole pockets have distinct sizes due to correlation effects. In particular, the innermost hole pocket is negligibly small. In the doping range of $0.1 < x < 0.2$, a Lifshitz transition occurs in which the inner cylinder-like hole Fermi surface shrinks into a line and then reemerges as a cylinder-like electron Fermi surface. This Lifshitz transition occurs at the same doping region where the experimental Tc is suddenly enhanced. Therefore, the abrupt Tc jump around x=0.15 is probably caused by such a Lifshitz transition. Our view is further supported by the previous angle-resolved photoemission spectroscopy (ARPES) experiments which demonstrated that a similar Lifshitz transition is accompanied by a large $T_c$ enhancement in the K-doped FeSe mono-layer superconductor. At very high doping levels (x>0.5), a new electron pocket with dominate $d_{xy}$ orbital character can be introduced at the Γ point in the overdoped region. This Lifshitz transition is recently observed in both K doped monolayer FeSe and surface K–doped $(Li_{0.8}Fe_{0.2}OH)FeSe$, but in these two cases the responses of Tc are opposite[21,23]. Our results suggest the latter Lifshitz transition is likely aroused by elevating of Se height in the overdoped region. However in the current system we find no signature of a Lifshitz transition induced Tc enhancement in the overdoped region.

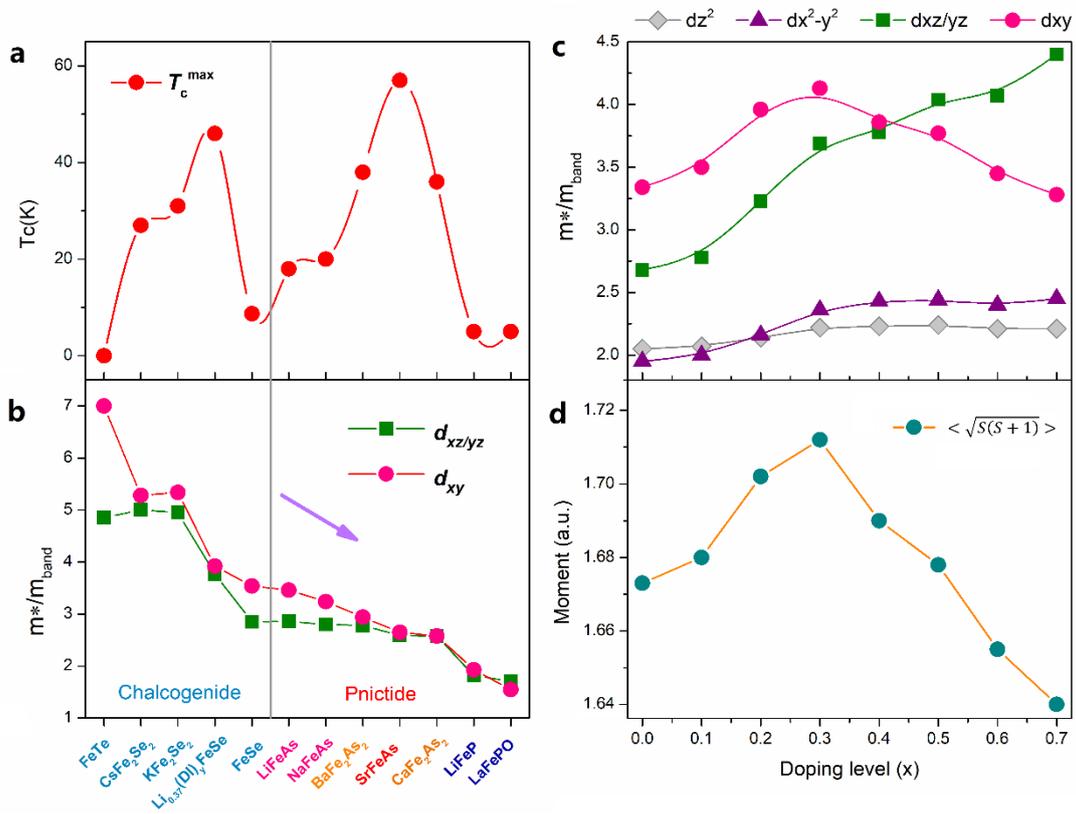

**Figure 5. Maximum Tc , mass enhancement and fluctuating local spin . a,** Maximum experimental Tc achieved by electron or hole doping near the parent compound in iron chalcogenides and iron pnictides. **b,** The DFT+DMFT calculated mass enhancement $m^*/m_{band}$ of the iron 3d xz, yz, and xy orbitals in the paramagnetic state [32]. Maximum Tc is achieved when the dxz/yz orbital and dxy orbital have almost the same, intermediate mass enhancement (~3 and ~4 respectively) in both iron chalcogenides and iron pnictides. Tc is smaller when there is larger orbital selectivity or differentiation among the Fe 3d t2g orbitals. **c,** Mass enhancements of Fe *3d* orbitals as a function

of doping in $Li_x(C_3N_2H_{10})_{0.37}FeSe$. The maximal Tc occurs in the doping region where there is a crossover of the mass enhancement of the dxz/yz orbital and the dxy orbital, further confirming the observation in **a** and **b**. **d,** Fluctuating local spin of Fe *3d* states versus doping in $Li_x(C_3N_2H_{10})_{0.37}FeSe$. The fluctuating local spin strongly correlates with the experimental superconducting transition temperature in the doping region 0.1<x<0.7 where nematic phase and magnetic ordering are absent, suggesting spin fluctuations is the driving force for the observed superconductivity in this family of materials.

In iron-based superconductors, due to the multi-band multi-orbital nature of their electronic structures, it is difficult to correlate the superconducting temperature with one variable. However, a comprehensive ARPES study[33] of various iron-based superconductors has shown that electron correlation strength plays the most important role in the superconductivity of iron-based superconductors. An analysis of the mass enhancement[32] and the maximum experimental Tc achieved by electron or hole doping near the parent compound in various iron chalcogenides and iron pnictides (see Fig. 5**a**,**b**) indicates that the highest Tc is achieved when both the dxz/yz orbital and the dxy orbital have almost the same mass enhancement, around 3 in iron pnictides and 5 in iron chalcogenides (4 if we include the data for $Li_{0.37}(C_3N_2H_{10})_{0.37}FeSe$). Tc is lower when there is larger orbital selectivity or differentiation among the Fe 3d t2g orbitals. This analysis suggests that reducing the electronic correlation strength difference between the dxz/yz and dxy orbitals helps to boost the superconducting temperature of iron-based superconductors. Therefore it is interesting to see how the mass enhancement evolves with doping in $Li_x(C_3N_2H_{10})_{0.37}FeSe$.

Previous DFT+DMFT calculations and model studies[32,34,35,36,37] in iron superconductors have shown that the electronic correlation strength of the Fe 3d electrons is sensitive to both the anion height and the Fe 3d occupancy (the doping level). The increase in Fe-Se distance and the Se height are expected to promote stronger electronic correlation at higher Li doping level. On the other hand, higher electron doping means higher Fe 3d occupancy, hence weaker electronic correlation. We plot the DFT+DMFT mass enhancement of Fe *3d* orbitals at different Li doping levels in Figure 5 **c**. In the whole doping range (0-0.7), mass enhancements of $e_g$ orbitals are around 2, whereas mass enhancements of $t_{2g}$ orbitals are much higher, with an average around 3.5. Due to the competing effects of charge doping and change of Se height, the mass enhancement of the *$d_{xz/yz}$* orbital increases with increasing Li doping levels monotonically whereas the mass enhancement of the *$d_{xy}$* orbital first increases and then decreases with Li doping with a maximum at x=0.3. Therefore, electron doping to the superconducting FeSe layer opens up a new avenue for fine-tuning the (relative) strength of mass enhancement or electronic correlation of the Fe *3d* electrons, which in turn can fine-tune the superconductivity. Indeed, ours results show that around the optimal doped region x ≈0.37 where the highest Tc=46 K is achieved, there is a crossover from *$d_{xy}$* to *$d_{xz/yz}$* being the most strongly correlated orbital. At the crossover, all the Fe *3d* $t_{2g}$ orbitals are equally correlated with mass enhancement of about 3.8. The fact that the crossover occurs around the optimal doping region further indicates that minimizing orbital selectivity or differentiation in electronic correlation strength of Fe 3d t2g orbitals, by means of tuning the crystal structure and carrier doping, can enhance superconductivity.

Spin fluctuation is another important quantity that is believed to be closely connected to superconductivity in iron-based superconductors [38]. We computed and shown in Fig. 5d the fluctuating local spin moment of Fe *3d* electrons as a function of Li doping level. In the experimental region 0.06≤ x ≤0.68, static nematic and magnetic orderings are absent according to our experiments. In this same region, the calculated fluctuating local spin moment and the experimental $T_c$ show a similar variation with Li doping. In particular, the calculated local fluctuating spin moment is maximized around the optimal doping. This suggests that the strong spin fluctuation in the $Li_x(C_3N_2H_{10})_{0.37}FeSe$ superconductors is likely the driving force for its high temperature superconductivity.

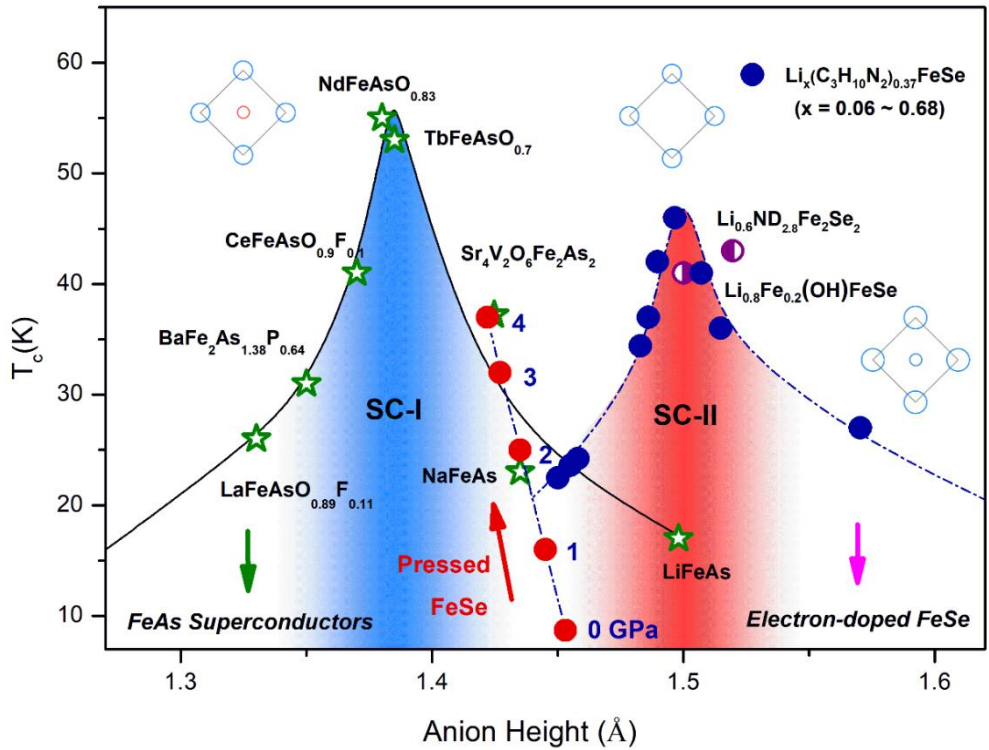

**Figure 6. Two superconducting zones in iron-based superconductors.** The superconducting temperature as a function of anion height in many iron-based superconductors is shown here. The Tc values were obtained from resistivity or susceptibility measurements. The iron pnictides, stoichiometric FeSe under pressure and electron doped FeSe-based samples show two different superconducting zones (SC-I and SC-II) separated by a critical anion height of around 1.45 Å. The SC-I superconducting region features optimal superconductivity near perfect Fe-anion tetrahedron with anion height ~1.38 Å, where the dxz/yz and dxy orbitals have almost the same intermediate mass enhancement around 3 as indicated in Fig.5 **a** and **b**. The maximum Tc ~55 K was found in $NdFeAsO_{0.83}$. The SC-II superconducting region has strongly distorted Fe-anion tetrahedron with large anion height. The maximum Tc is found to be 46K at anion height of ~1.50 Å, where the dxz/yz and dxy orbitals have almost the same intermediate mass enhancement around 4 as indicated in Fig.5 **a**, **b** and **c**. Error bars are within the size of the symbols.

**Discussions**

Previously, it was found empirically that the superconducting temperature is maximized around a unique anion height across many families of iron-based superconductors[32,39]. In Figure 6, we plot the anion height dependence of superconducting temperature in $Li_x(C_3N_2H_{10})_{0.37}FeSe$ as well as other Fe-based superconductors reported previously[2,40,41,42,43,44,45,46,47]. It is found that in iron arsenides, the maximum Tc is realized in $NdFeAsO_{0.83}$ with an ideal $FeAs_4$ tetrahedron, and the optimal anion height is around 1.38 Å. Similarly, pressurized FeSe also follow the same role, as the high pressure reduced the structural distortion in pristine FeSe by decreasing $h_{Se}$ from 1.45 Å toward 1.38 Å, the Tc is drastically increased by a factor of four. Moreover, the static antiferromagnetic order and large hole pockets that widely observed in iron pnictides appeared in FeSe under pressure[47,48,49], also indicating pressurized FeSe belong to the same superconducting zone host by iron pnictides (SC-I). However, as shown in Figure 6, the observed evolution of anion height of FeSe layer under electron doping deviate monotonically away from the superconducting zone of high-Tc iron arsenides and pressurized FeSe, with Se-Fe-Se angle significantly reduces to 100.9° and $h_{Se}$ mounts up to 1.57 Å in the over-doped regime. The effect of electron-doping in FeSe hence functions like a 'negative' pressure, pushing the system into a new superconducting zone. As demonstrated in our theoretical calculations, a combined effect of elevating anion height and carrier doping leads to successive Lifshitz transitions and two unique electronic structures with only electron pockets, completely different from the ground states in iron pnictides and pressurized FeSe. The fact that electron-doped FeSe belongs to a new superconducting zone that remote from the known superconducting zone host by pnictides and pressurized FeSe well explained the origin of the extraordinary behaviors emerged in electron-doped FeSe.

The existence of two superconducting zones correlated with anion height in iron based superconductors also suggest there are different ways to optimize Tc by tune the structure parameters and carrier concentration. In most iron arsenides (including pressurized FeSe), a perfect FeAs4 (FeSe4) tetrahedron minimizes the orbital differentiation of the Fe *3d* $t_{2g}$ orbitals and an optimal As height around 1.38 Å give rises to an intermediate mass enhancement ($m^*/m_{band}$ ~3) of the $t_{2g}$ orbitals. Although in $Li_x(C_3N_2H_{10})_{0.37}FeSe$, the FeSe4 tetrahedron is far away from perfect tetrahedron near optimal Li doping, the coupled Se height and carrier concentration around the optimal doping x~0.37 lead to the minimal orbital differentiation of the Fe *3d* $t_{2g}$ orbitals with an intermediate mass enhancement ($m^*/m_{band}$ ~3.8). Therefore, by means of tuning the crystal structure and carrier concentration, a common strategy to maximize Tc in both iron-arsenide and iron selenide superconductors is to minimize the Fe 3d $t_{2g}$ orbital differentiation and couple with intermediate electronic correlation to promote large spin fluctuations.

Our results also support a natural origin of the emergence of a high Tc superconducting phase in electron-doped iron selenides under high pressure, including $A_xFe_{2-y}Se_2$ (48K, 12Gpa)[11], LiOHFeSe (52K, 12.5Gpa)[50] and most recently in $Li_{0.4}(NH_3)_yFeSe$ (55K, 11.5Gpa)[51]. As shown in Figure 6, high pressure is known to decrease the anion height of FeSe superconductors. For electron-doped FeSe superconductors, the minimal orbital differentiation of Fe *3d* t2g orbitals is realized at ambient pressure with a large Se height ($h_{Se}$) of 1.51Å. While $h_{Se}$ decreases from 1.51 Å to ~1.45 Å under pressure, the electron doped iron selenides are leaving the SC-II zone, therefore their Tc first decrease under pressure (Fig.6). Upon increasing pressure, $h_{Se}$ is further reduced towards 1.38 Å, they then enter the SC-I zone and the orbital differentiation of Fe *3d* t2g orbitals is gradually reduced as the FeSe$_4$ tetrahedron approaches the ideal tetrahedron where the optimal Tc was achieved in iron arsenide superconductors. As a result, a new superconducting phase with even higher Tc emerges, with the maximum Tc =55 K equal to the optimized Tc of iron arsenides. Future detailed structural analysis of electron doped iron selenides under pressure and DFT+DMFT calculations are required to confirm this picture.

In summary, we realized continuous electron doping in a new family of bulk FeSe-based superconductor-Li$_x$(C$_3$N$_2$H$_{10}$)$_{0.37}$FeSe, which features a continuous superconducting dome harboring Lifshitz transitions within the wide range of 0.06⩽x⩽0.68. We demonstrate that with increasing electron doping, the anion height of FeSe layers almost linearly deviates away from the optimal anion height value of iron pnictides and pressurized FeSe, and falls into a new superconducting zone well separated from those superconductors. Our DFT+DMFT calculations reveal the key roles of anion height and carrier concentration in determining the electronic structures and the strong orbital-selective correlation effects unique in electron-doped FeSe. By summarizing the correlation of Tc's and the mass enhancement of t$_{2g}$ orbitals of iron pnictides, pressurized FeSe and the electron doped FeSe-based superconductors, we find a common theme to optimize Tc by means of doping and pressure: minimizing the Fe *3d* t$_{2g}$ orbital differentiation coupled with intermediate electronic correlation by tuning the anion height and carrier concentration. The new phase diagram provides a unified picture of electron-doped and pressurized FeSe superconductors and iron-arsenide superconductors.

# References

1. Igor I. Mazin, Superconductivity gets an iron boost. *Nature* **464** (2010).

2. Kamihara, Y., Watanabe, T., Hirano, M. & Hosono, H. Iron-based layered superconductor La[$O_{1-x}F_x$]FeAs (x = 0.05−0.12) with Tc = 26 K. *J. Am. Chem. Soc.* **130**, 3296 (2008).

3. Lee, P. A, Nagaosa, N. & Wen, X. G. Doping a Mott insulator: physics of high-temperature superconductivity. *Rev. Mod. Phys.* **78**, 17–85 (2006).

4. Stewart, G. R. Superconductivity in iron compounds. *Rev. Mod. Phys.* **83**, 1589-1652 (2011).

5. Paglione J. & Greene R. L. High-temperature superconductivity in iron-based materials. *Nat. Phys.* **6**, 645-658 (2010).

6. S. Medvedev, *et al*. Electronic and magnetic phase diagram of $\beta$-$Fe_{1.01}$Se with superconductivity at 36.7 K under pressure. *Nat. Mater.* **8**, 2491 (2009).

7. Alireza, P. L. *et al.* Superconductivity up to 29 K in $SrFe_2As_2$ and $BaFe_2As_2$ at high pressures. *J. Phys. Condens. Matter* **21**, 012208 (2008).

8. Simon A. J. Kimber, *et al*. Similarities between structural distortions under pressure and chemical doping in superconducting $BaFe_2As_2$. *Nat. Mater.* **8**, 2443 (2009).

9. Qian, T. *et al.* Absence of a hole like fermi surface for the Iron-Based $K_{0.8}Fe_{1.7}Se_2$ superconductor revealed by angle-resolved photoemission spectroscopy. *Phys. Rev. Lett.* **106**, 187001 (2011).

10. Hirschfeld, P. J., Korshunov, M. M. & Mazin, I. I. Gap symmetry and structure of Fe-based superconductors. *Rep. Prog. Phys.* **74**, 124508 (2011).

11. Sun, L. L. *et al.* Re-emerging superconductivity at 48 kelvin in iron chalcogenides. *Nature* **483,** 67–69 (2012).

12. Sun, J. P. *et al.* Reemergence of high-Tc superconductivity in the ($Li_{1-x}Fe_x$)$OHFe_{1-y}Se$ under high pressure. *Nat. Comm*. **9**, 380 (2018).

13. Zhang, W. H. *et al*. Direct observation of high-temperature superconductivity in One-Unit-Cell FeSe films. *Chin. Phys. Lett.* **31**, 017401 (2014).

14. He, S. L. *et al.* Phase diagram and electronic indication of high-temperature superconductivity at 65K in single-layer FeSe films. *Nat. Mater.* **12**, 605-610 (2013).

15. Guo, J. G. *et al*. Superconductivity in the iron selenide $K_xFe_2Se_2$(0≤x≤1.0). *Phys. Rev. B* **82**, 189520 (2010).

16. Wang, A. F. *et al.* Superconductivity at 32K in single-crystalline $Rb_xFe_{2-y}Se_2$. *Phys. Rev. B* **83**, 060512(R) (2011).

17. Krzton-Maziopa, A. *et al.* Synthesis and crystal growth of $Cs_{0.8}(FeSe_{0.98})_2$: A new iron-based

superconductor with $T_C$=27 K. *J. Phys. Condens. Matter* **23**, 052203 (2011).

18. Lu, X. F. *et al.* Coexistence of superconductivity and antiferromagnetism in $(Li_{0.8}Fe_{0.2})OHFeSe$. *Nat. Mater.* **15**, 4155 (2014).

19. Y. Miyata. *et al*. High-temperature superconductivity in potassium coated multilayer FeSe thin films. *Nat. Mater.* **14**, 775 (2015).

20. Lei, B. *et al*. Evolution of high-temperature superconductivity from a low-Tc phase tuned by carrier concentration in FeSe thin flakes. *Phys. Rev. Lett.* **116**, 077002 (2016).

21. Shi, X. *et al*. Enhanced superconductivity accompanying a lifshitz transition in electron-doped FeSe monolayer. *Nat. Comm*. **8**, 14988 (2017).

22. Wen, C. H. P. *et al*. Anomalous correlation effects and unique phase diagram of electron-doped FeSe revealed by photoemission spectroscopy. *Nat. Comm*. **7**, 10840 (2016).

23. Ren, M. Q. *et al.* Superconductivity across lifshitz transition and anomalous insulating state in surface K-doped $(Li_{0.8}Fe_{0.2})OHFeSe$. *Science Adv.* **3**, e1603228 (2017).

24. Huang, D. *et al.* Revealing the empty-state electronic structure of single-unit-cell $FeSe/SrTiO_3$. *Phys. Rev. Lett.* **115**, 017002 (2015).

25. Matthew, Burrard-Lucas. Enhancement of the superconducting transition temperature of FeSe by intercalation of a molecular spacer layer. *Nat. Mater.* **12**, 3464 (2013).

26. Sun, H. L. *et al*. Soft Chemical control of superconductivity in lithium iron selenide hydroxides $Li_{1-x}Fe_x(OH)Fe_{1-y}Se$. *Inorg. Chem.* **54**, 1958-1964 (2015)

27. Zhao, J. *et al*. Structural and magnetic phase diagram of $CeFeAsO_{1-x}F_x$ and its relation to high-temperature superconductivity. *Nat. Mater.* **7**, 953–959 (2008).

28. Rey, R. I. *et al*. Quasi-2D behavior of 112-type iron-based superconductors. *Supercond. Sci. Technol.* **26**, 055004 (2013).

29. Hikami, S. Larkin, A.I. Magnetoresistance of high temperature superconductors. *Mod. Phys. Lett. B* **2**, 693 (1988).

30. Huang, D. *et al.* Revealing the empty-state electronic structure of single-unit-cell $FeSe/SrTiO_3$. *Phys. Rev. Lett.* **115**, 017002 (2015).

31. Haule, K. Yee, C-H. & Kim, K. Dynamical mean-field theory within the full-potential methods: electronic structure of Ce-115 materials. *Phys. Rev. B* **81**, 195107 (2010).

32. Yin, Z. P. Haule, K. & Kotliar, G. Kinetic frustration and the nature of the magnetic and paramagnetic states in iron pnictides and iron chalcogenides. *Nat. Mater.* **10**, 932 (2011).

33. Ye, Z. R. *et al.* Extraordinary Doping Effects on Quasiparticle Scattering and Bandwidth in Iron-Based Superconductors *Phys. Rev. X.* **4**, 031041 (2014).

34. Mathias, Duckheim. & Piet, W. Brouwer. Andreev reflection from noncentrosymmetric superconductors and Majorana bound-state generation in half-metallic ferromagnets. *Phys. Rev. B* **83**, 054513 (2012)

35. Takahiro, Misawa. Kazuma, Nakamura. & Masatoshi, Imada. Ab Initio Evidence for Strong Correlation Associated with Mott Proximity in Iron-Based Superconductors. *Phys. Rev. Lett.* **108**, 177007 (2012).

36. Luca, de' Medici. Gianluca, Giovannetti. & Massimo, Capone. Selective Mott Physics as a Key to Iron Superconductors. *Phys. Rev. Lett.* **112**, 177001 (2014).

37. Yin, Z. P. Haule, K. & Kotliar, G. Fractional power-law behavior and its origin in iron-chalcogenide and ruthenate superconductors: Insights from first-principles calculations. *Phys. Rev. B* **86**, 195141 (2012).

38. Dai, Pengcheng. Antiferromagnetic order and spin dynamics in iron-based superconductors. *Rev. Mod. Phys.* **87**, 855 (2015).

39. Mizuguchi, Y. *et al*. Anion height dependence of Tc for the Fe-based superconductor. *Supercond. Sci. Technol.* **23**, 054013 (2010).

40. Alex, J. Corkett *et al.* Control of the superconducting properties of $Sr_{2-x}Ca_xVO_3FeAs$ through isovalent substitution. *Journal of Solid State Chemistry* **216,** (2014).

41. Chen, G. F. *et al*. Superconductivity at 41 K and its competition with spin-density-wave instability in layered $CeO_{1-x}F_xFeAs$. *Phys. Rev. Lett.* **100**, 247002 (2008).

42. Lee, C. H. *et al*. Effect of structural parameters on superconductivity in fluorine-free $LnFeAsO_{1-y}$ (Ln = La, Nd). *J. Phys. Soc. Japan* **77**, 083704 (2008).

43. Allred, J. M. *et al*. Coincident structural and magnetic order in $BaFe_2(As_{1-x}P_x)_2$ revealed by high-resolution neutron diffraction. *Phys. Rev. B* **90**, 104513 (2014).

44. Pitcher, M. J. *et al.* Structure and superconductivity of LiFeAs. *Chem. Comm.* **44**, 5918-5920 (2008).

45. Dinah, R. *et al.* Structure, antiferromagnetism and superconductivity of the layered iron arsenide NaFeAs. *Chem. Comm.* **16**, 2189 (2009).

46. Kiichi M. *et al.* Superconductivity above 50K in $LnFeAsO_{1-y}$ (Ln=Nd, Sm, Gd, Tb, and Dy) Synthesized by High-Pressure Technique. *J. Phys. Soc. Japan* **78**, 034712 (2009).

47. Terashima, T. *et al.* Pressure-induced antiferromagnetic transition and phase diagram in FeSe. *J. Phys. Soc. Jpn.* **84**, 063701 (2015).

48. Sun, J. P. *et al.* Dome-shaped magnetic order competing with high-temperature superconductivity at high pressures in FeSe. *Nat. Comm.* **7**, 12146 (2016).

49. Sun, J. P. *et al.* High-Tc Superconductivity in FeSe at High Pressure: Dominant Hole Carriers and Enhanced Spin Fluctuations. *Phys. Rev. Lett.* **118**, 147004 (2017).

50. Sun, J. P. *et al.* Reemergence of high-Tc superconductivity in the $(Li_{1-x}Fe_x)OHFe_{1-y}Se$ under high pressure. *Nat. Comm*. **9**, 380 (2018).

51. Shahi, P. *et al*. High-Tc superconductivity up to 55 K under high pressure in a heavily electron doped $Li_{0.36}(NH_3)_yFe_2Se_2$ single crystal. *Phys. Rev. B* **97**, 020508 (2018).

**Acknowledgements**

This work was supported by the National Natural Science Foundation of China (Grant No. 51472266, 11674030, 11704034, 51772323,51532010), the Fundamental Research Funds for the Central Universities (Grant No.310421113) and the National Key Research and Development Program of China through Contract No. 2016YFA0302300,2016YFA0300301, the National Youth Thousand-Talents Program of China, and the start-up funding of Beijing Normal University. The calculations used high performance computing clusters of Beijing Normal University in Zhuhai and the National Supercomputer Center in Guangzhou.

**Author contributions**

S.F.J. and Z.P.Y. conceived and led this work. R.J.S. and Y. Q. contributed equally to this work. R. J. S. synthesized the materials and performed the resistance and magnetic susceptibility measurements. Y.Q. and Z.P.Y. did the DFT+DMFT calculations. Q. H. and H. W. performed the neutron diffraction experiments. S.F.J. and Z.P.Y. wrote the paper with input from R.J.S. and Y.Q. All authors discussed the results and commented on the manuscript. X. L. C. and L. G. directed the whole project.

**Additional information**

Supplementary information is available in the online version of the paper. Correspondence and requests for materials should be addressed to S.F.J., Z.P.Y., L. G. or X.L.C.

**Competing financial interests**

The authors declare no competing financial interests.